# Some insights into the behaviour of millimetre wave spectrum on key 5G cellular KPI's

Mythri Hunukumbure[1], Yue Wang[1], Miurel Tercero[2], Javier Lorca[3]
[1]Samsung Electronics R&D Institute UK, Staines, Middlesex TW18 4QE, UK
[2]Ericcson AB, Stockholm, Sweden
[3]Telefónica I+D – GCTO, C/ Zurbarán 12, 28010 Madrid, Spain

[1]{mythri.h, yue2.wang}@samsung.com, [2]miurel.i.tercero@ericsson.com, [3]franciscojavier.lorcahernando@telefonica.com

*Abstract*—This invited paper discusses the challenging task of assessing the millmeter-wave spectrum for the use of 5G mobile systems. 3 key KPIs related to mobile systems, coverage, capacity and mobility are individually assessed and quantified with a simplistic figure of merit (FoM). The coverage and capacity KPIs are assessed through simulations while an analytical approach is developed for mobility. The key message is that the FoM degrades with the higher as the frequencies increase in this range. This indicates that there will be more challenges in deploying millimeter wave systems in higher frequencies, but these are not insurmountable. Many of these challenges are addressed in the mmMAGIC project and possible solutions are derived.

*Keywords—5G, coverage, capacity, mobility, millimeter wave, spectrum assessment*

## I. Introduction

The millimeter wave (mm-wave) spectrum commonly referred to as between 6 and 100 GHz and the associated radio technologies are recognized as key components for the up-coming 5G communication standards. 3GPP completing their initial work on the above 6GHz channel model and starting the study item on New Radio (NR) [1], the Federal Communications Commission's (FCC) identification of 28 GHz and 37-39 GHz bands for licensed mobile use [2] and the European Commission naming a European 5G pioneer band at 26 GHz [3] are some of the major developments in this area in the past year. Within this backdrop of rapid developments, the technical work carried out by the mmMAGIC project [4], i.e. to design and develop key concepts and components of a mm-wave RAT (Radio Access Technology), carries added significance. This paper details some of the main work conducted in the project in the area of mm-wave spectrum assessment.

In this paper, we consider the performance of 3 key KPIs needed for effective 5G cellular communications - coverage, capacity and mobility – in the mm-wave spectrum. These KPIs cover the essential elements for the provision of eMBB (enhanced Mobile Broadband), widely seen as the first use case that will be addressed by 5G. For coverage and capacity, we analyze the performance through simulations carried out on specific mm-wave frequencies, but we also highlight the trends coming out of this analysis. We focus on some of the key mm-wave bands identified in the WRC-15 (World Radio-communication Congress, 2015), for further suitability study on mobile applications [5]. For the study on mobility, we develop an analytical approach and study the trends with this.

The remainder of this paper is organized as follows. In section II, we present the detailed assessment methodology and also discuss some limitations in this approach. Section III is devoted to presenting the results and discussing the trends in the behavior of the coverage KPI. The capacity and mobility KPIs are assessed and discussed in sections IV and V respectively. The conclusions from this work are presented in section VI.

## II. Assesment Methodology

There are multiple parameters which interact in potential mobile communication systems in the mm-wave spectrum. One of the main attractions in mm-wave is the availability of very wide bandwidths, which can potentially compensate the negative impacts of excessive path loss in these frequencies. This is particularly true in higher GHz frequencies, where the path loss is more severe, yet there are very wide bandwidths potentially available. By increasing the number of antenna elements for higher beam-forming gain, the path loss can be countered as well. As the spacing of antenna elements gets physically smaller in higher frequencies, the antenna increment can be achieved without a significant increase in the array sizes. The cell sizes can also be adjusted to provide the targeted coverage and capacity.

Due to the complex interactions of these multiple parameters, we propose to conduct this analysis as two strands. Some of the parameters will be fixed in each strand and this allows the variation of other parameters across the frequency range to be analysed with more clarity. The two strands will be applied across all 3 KPIs, so a fairer comparison on the impacts on each KPI can be obtained. Both the analyses will be comparative studies, i.e. we study the relative impact of increasing carrier frequency has on these selected KPIs as referenced to a carrier frequency at 6GHz. Thus $f_{ref}$=6GHz for this study. The two strands can be listed as follows;

- Strand 1: Fix the system bandwidth to around 500MHz and study the impact of incrementing number of antenna elements. The antenna numbers

will be incremented in line with $\left(\frac{f_c}{f_{ref}}\right)^2$ to counter for the incrementing path loss. Here $f_c$ is the carrier frequency under consideration.

- Strand 2: Fix the number of antenna elements ($N_{ant}$=32) across the frequency range and study the impact of incrementing the system bandwidth. The bandwidth is incremented in line with the values obtained from the coverage KPI analysis.

These incrementing antenna numbers and bandwidth in this arbitrary manner will not be a practical option in many situations. Both these increments will have resulting hardware complexities, which are not considered in the study. Here we consider quite a hypothetical situation, to isolate the impact of mm-wave spectrum on the KPIs, for specific outdoor or outdoor to Indoor (in case of capacity KPI).

Two of the KPIs, coverage and capacity, are assessed with the aid of multi-cell simulations. The mobility analysis is based on theoretical quantitative analysis on some related parameters. The coverage and capacity analyses are limited to spot frequency values in some of the parameters. To simplify the quantification of the assessments, we derive a Figure of Merit (FoM) in the range of [0 10] for each of the KPIs.

## III. COVERAGE KPI ANALYSIS

The coverage KPI in this study is assessed in terms of the cell sizes needed to achieve a certain cell edge data rate. We run multi-cell simulations with multiple users and build up a data set on the data rates achieved by the cell edge users, in light of the path loss and shadowing of the signal (S), noise accumulated (N) and the interference (I) from the other users. The cell data rate ($R_{CE}$) is calculated using the fundamental Shannon equation as follows;

$$R_{CE} = B \cdot log_2\left(1 + \frac{S}{N+I}\right) \quad (1)$$

where the additional term B is the bandwidth occupied by each user. The cell edge data rates for multiple users are captured and the 95th percentile of the cdf (cumulative distribution function) is taken as the representative data rate.

The noise power is considered to be additive white Gaussian and proportionally incremented with the bandwidth (N=kTB). The signal power is derived from the following relationship;

$$S = P + G_{BF} - PL \quad (2)$$

where P is the transmit power, $G_{BF}$ is the cumulative antenna gains (transmitter and receiver sides) and PL is the path loss. The path loss is a critical factor here and reliable models are still emerging for this mm-wave frequency range, for cellular applications. We utilize the recent path loss models from NYU (work by Rappaport et.al.) [6], where they propose path loss models derived from extensive measurements for the 28GHz, 38GHz and 73GHz carrier frequencies. The simulations need fixed, spot carrier frequencies and we select the above 3 frequencies covering the low, mid and high GHz ranges. Additionally, we select the $f_{ref}$=6GHz as the reference simulation and use the WINNER II path loss model for this frequency.

The bench mark for coverage analysis is set at achieving 100Mbps data rate at the cell edge and it is statistically analysed with data rates achieved by multiple cell edge users.

### A. Coverage KPI – Strand 1 assessment:

As noted above, in this strand we fix the bandwidth and study the number of cells needed to provide the 100Mbps cell edge data rate. The antenna numbers are incremented accordingly. The simulations however need the antenna numbers (n) to be a power of 2. The lowest carrier frequency applicable in the simulations with n=4 is 10GHz, so we use it as a shifted reference value in this work. The simulations are run for relatively fixed bandwidth values in the range of 500-600MHz. The results summary for this analysis is listed below in Table 1.

Table 1: Coverage KPI analysis – Strand 1

| Carrier frq (GHz) | 10 | 28 | 38 | 73 |
|---|---|---|---|---|
| No of Antennas | 4 | 32 | 64 | 256 |
| BW (MHz) | 500 | 500 | 500 | 600 |
| Cell radius (m) | 250 ($R_{ref}$) | 250 | 250 | 100 |
| Coverage - No. of cells ($R_{ref}$/R)$^2$ | 1 | 1 | 1 | 6.25 |
| Figure of Merit | 10 | 10 | 10 | 1.6 |

The simulation results indicate that up to the mid GHz range in the mm-wave spectrum, the path loss can be effectively compensated by the increment in the antenna numbers, roughly at the rate of square of the frequency increment. There are other factors considered in the simulator and the path loss models, like the increasing probability of shadowing when the beams get narrower. These effects make the cell sizes much smaller for the 73GHz carrier to achieve the 100Mbps cell edge data rate and hence the Figure of Merit (FoM).

### B. Coverage KPI – Strand 2 assessment:

In strand 2 of coverage assessment, we fix the number of antennas to 32 across the frequency range and analyse the cell sizes needed to achieve the 100Mbps cell edge data rate, while changing the bandwidth.

In interpreting simulation results, we have looked to minimize the cell radius, while allowing the bandwidth increment. However, when moving onto higher frequencies, only similar or higher bandwidths to the lower spot frequencies were considered, to be in line with the general trend of larger bandwidth availability for higher carrier frequencies. The results are reported in Table 2 below.

Table 2: Coverage KPI analysis – Strand 2

| Carrier Freq (GHz) | 6 | 28 | 38 | 73 |
|---|---|---|---|---|
| BW (MHz) | 400 | 400 | 500 | 1400 |
| Cell radius (m) | 200 ($R_{ref}$) | 150 | 100 | 50 |

| Coverage - No. of cells ($R_{ref}/R)^2$ | 1 | 1.77 | 4 | 16 |
|---|---|---|---|---|
| Figure of Merit | 10 | 5.625 | 2.5 | 0.0625 |

The bandwidth expansion has a direct impact on the link budget as the noise power increments and thus impacts the data rate as in (1). The may be overcome by having component carriers as in carrier aggregation, but we have not factored this in this anaylsis. The cell sizes successively become smaller and hence the number of cells required to achieve coverage in a unit area increases. We have quantified the FoM value for this strand 2 accordingly.

## IV. Capacity KPI Analysis

The capacity KPI is also assessed with the aid of multi-cell simulations. As with the coverage analysis, only spot values are available in certain parameters, due to the complexity and the time consumption in the simulations.

In this scenario the macro base station will be serving indoor users, from a rooftop of an adjacent building. The outdoor to indoor propagation model consider FSPL, diffraction loss, indoor loss and body loss. For O2I (outdoor to indoor case), the average data rate is computed from the users that are indoor in the border of the cell range. The indoor depths of the user vary from 0.5 to 10 m.

For scenario1 the bandwidth is assumed to be constant to 100MHz and for scenario 2 it is assumed to be increasing with the frequency according to 5% the central frequency. The directivity of the transmission enables less interference between the links. In the outdoor network in mm-wave bands, the highly directional links can be modelled as "pseudo-wired". As a first assumption, we then consider negligible the inter cell interference between non-adjacent links [7-8]. In mm-wave systems the thermal noise dominates interference: highly directional transmissions used in mm-wave systems combined with short cell radii result in links that are in relatively high SINR with little interference. As a first assumption we then can assume that there is no intra-cell interference. We have considered the case where a user is in good propagation conditions (i.e. lower building penetration loss) in this analysis.

Then the received signal (Rx) can be computed as:

$$Rx = EIRP + G_{Rx} - PL \quad (3)$$

The average data rate is dependent on:
$$DataRate_{Avr} = \left[\rho BW \log_2\left(1 + \frac{Rx}{Noise}\right)\right]_{Avr} \quad (4)$$

The area capacity is computed as:
$$Area\ Capacity = \left(\frac{cells}{Km^2}\right) * \left(\frac{Ues}{cell}\right) * DataRate_{Avr} \quad (5)$$

A particular feature that was considered in the capacity analysis is that the number of antennas cannot increase infinitely. Regarding antenna assumption, the area (A) of the antenna is assumed to be 1x0.1 m2, and it is kept constant for all frequencies. However, the antenna gain is computed as (where c is the speed of light):

$$G = 4\pi(f_{GHz})^2/c^2 \quad (6)$$

This means that the antenna gain will increase with frequency by keeping the same antenna area. However, it is not realistic to assume infinite increment of gain, thus a maximum 50dBi transmitter/ receiver combine beamforming gain is considered.

The antenna gain variation across the carrier frequency range is shown below.

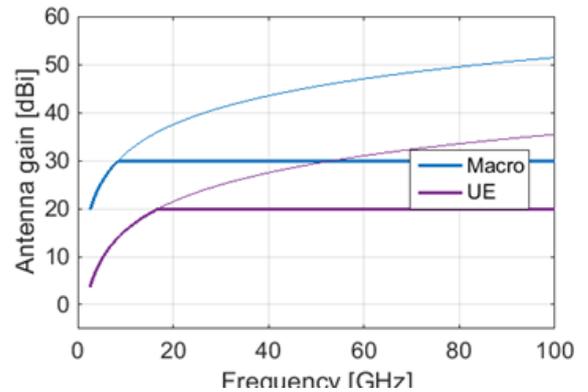

Figure 1: Antenna gain variations considered in the capacity KPI analysis

This antenna gain consideration puts some restrictions in the way the capacity KPI can be analysed in strand 1 and 2. The strand 1, where the number of antenna elements (and hence the antenna gain) increment with the carrier frequency, can be only analysed for 6-20 GHz frequency range. Similarly for strand 2, where the antenna gain is constant, only the frequency range 20-100 GHz can be employed for the analysis.

### A. Capacity KPI – Strand 1 assessment:

The capacity assessment takes the area capacity values generated by the simulations for the 6-20GHz range, where the antenna gains show increment. The system bandwidth is fixed to 500MHz and the inter site distance (ISD) to 300m. These fixed values resemble similar trends for the coverage analysis (strand 1) in Table 3.

Table 3: Capacity KPI analysis – Strand 1

| Carrier frq (GHz) | 6 ($f_{ref}$) | 10 | 20 |
|---|---|---|---|
| Area capacity (Gbps/km²) | 297 | 284 | 185 |
| Normalized to $f_{ref}$ | 1 | 0.95 | 0.62 |
| Figure of Merit | 10 | 9.5 | 6.2 |

The results indicate that the capacity KPI show a gradual decrement as per the incrementing carrier frequency. This is in line with the findings that coverage can be largely maintained with the higher number of antennas in this frequency range.

Conversely, the capacity KPI can be maintained with increasing the number of cells gradually in this range, but it will then impact the coverage KPI. These trends further show the inter-relationships amongst the KPIs.

*B. Capacity KPI – Strand 2 assessment:*

For the strand 2, the antenna gains are fixed and the BW and cell sizes are varied in order to achieve capacity. The cell radius and the bandwidth values have been aligned with the coverage KPI analysis.

Table 4: Capacity KPI analysis – Strand 2

| Carrier frq (GHz) | 20 ($f_{ref}$) | 28 | 38 | 73 |
|---|---|---|---|---|
| System BW (MHz) | 400 | 400 | 500 | 1400 |
| Cell radius (m) | 150 | 150 | 100 | 50 |
| Area capacity (Gbps/km$^2$) | 212 | 121 | 182 | 84 |
| Normalized to $f_{ref}$ | 1 | 0.57 | 0.86 | 0.40 |
| Figure of Merit | 10 | 5.71 | 8.58 | 3.96 |

As the simulation conditions make 20GHz the starting frequency for this analysis, we normalize the achieved area capacity values at this point and derive the suitability scaling. The values degrade gradually in the mid GHz range and then sharply in the High GHz range. This is despite compensating for the wider bandwidths and the smaller cell sizes available in the high GHz range. There is a slight anomaly that the capacity values have increased from 28 GHz to 38GHz, but this can be attributed to the coarse step changes in cell sizes (150m to 100m) as needed in coverage KPI analysis.

## V. MOBILITY KPI ANALYSIS

The two strands described above are analyzed for mobility assessment across frequency. Both cases require proper evaluation of the impact of frequency on mobility. Three phenomena are identified that may have significant impact: tracking effectiveness, Doppler spread, and channel coherence time. The dependence with frequency of these three effects is separately analyzed; an overall figure of merit is then obtained that provides the suitability of frequencies between 6 and 100 GHz in terms of mobility support; and application to the above two strands is highlighted in practical terms.

*A. Impact of tracking effectiveness*

Free-space attenuation increases with frequency following the quadratic dependency 20log(f), therefore beamforming gain will have to follow the same dependency with frequency. Gains in planar arrays are proportional to the product of the number of antennas in the H and V directions ($N_H$ and $N_V$ respectively), thus $G_{bf} \sim 10 \times \log(N_H N_V)$. If the start and end frequencies under analysis are denoted by $f_{min}$ and $f_{max}$ respectively, and to ideally overcome the increased pathloss, the following relation must be fulfilled:

$$20 log \frac{f_{max}}{f_{min}} = 10 log \frac{(N_H N_V)_{max}}{(N_H N_V)_{min}} \quad (7)$$

As an example, going from 10 GHz to 100 GHz represents 20 dB increase in path loss, therefore the total number of antennas will have to grow by a factor 100x (or 10x in each H, V directions).

Increased beamforming gain has a negative impact on device mobility, as the resulting beamwidth decreases accordingly which makes it more challenging to track the users. Neglecting the effect of the individual radiation patterns, the half-power beamwidth (HPBW) at the H and V planes is inversely proportional to the corresponding number of antennas:

$$\theta_{3dB,H} \cong \frac{k}{N_H}, \theta_{3dB,V} \cong \frac{k}{N_V} \quad (8)$$

If the users are concentrated in the same V plane then the tracking effectiveness would only depend on the horizontal HPBW. However it is unlikely that users remain in the same V plane if hotspots with reduced dimensions are targeted, where the distances between users and antennas can be small. If we define (for simplicity) the tracking effectiveness (TE) as the product of the half-power beamwidths in H and V, we have:

$$TA = \theta_{3dB,H} \cdot \theta_{3dB,V} \sim \frac{k'}{N_H N_V} \quad (9)$$

As a result, the relation between the defined tracking effectiveness at the maximum and minimum frequencies of analysis will be given by:

$$\frac{(TA)_{max}}{(TA)_{min}} = \frac{(N_H N_V)_{max}}{(N_H N_V)_{min}} = \left(\frac{f_{max}}{f_{min}}\right)^2 \quad (10)$$

In this sense, 100 GHz is 100 times worse than 10 GHz in terms of tracking effectiveness. It has to be emphasized that the above analysis considers antennas with ideally constant patterns, which may be somewhat unrealistic.

*B. Impact of Doppler spread*

Another factor that gets strong importance at higher frequencies is the Doppler spread $f_D$, defined as:

$$f_D = f_c \frac{v}{c} \quad (11)$$

where $f_c$ is the carrier frequency, $v$ is the user speed and $c$ is the speed of light. Therefore,

$$\frac{(f_D)_{max}}{(f_D)_{min}} = \frac{f_{max}}{f_{min}} \quad (12)$$

However, the real effects of Doppler spread in multicarrier systems (like OFDM and its variants) depends on the subcarrier width, which could scale with frequency so as to make the system robust to Doppler shifts even at very high frequencies. If this is the case then it would be complicated to assess the real effects of Doppler spread on the system.

Increasing the subcarrier width in multicarrier systems has the positive effect of reducing symbol length, but this of course has deep implications on the resulting numerology, which is unlikely to be allowed to change as flexibly as desired. One possible way forward could be to envisage a number of discrete alternatives for the numerology, preferably observing submultiple relationships, at targeted frequency points. Robustness to Doppler spread would then be maintained at those discrete frequencies.

C. *Impact of channel coherence time*

There is an additional negative effect linked to Doppler spread, namely the channel coherence time $T_c$, which follows the inverse relation $T_c \sim 1/f_D$. A shorter coherence time with frequency translates into poorer ability of the system to track channel variations (through link adaptation mechanisms and/or retransmissions). Linear dependency with frequency can therefore be stated also for $T_c$:

$$\frac{(T_c)_{max}}{(T_c)_{min}} = \frac{f_{max}}{f_{min}} \quad (12)$$

D. *Overall figure of merit*

The combination of the above three elements could lead to the definition of a figure of merit (e.g. between 0 and 10) that represents the inherent support of mobility at the different frequencies. The impact of frequency on tracking effectiveness could be assessed by taking the logarithm of the above defined tracking effectiveness, thus leading to a linear dependency with frequency of the type:

$$Mark_{TA} = MarkRef_{TA} - m_{TA} \cdot log\left(\frac{f}{f_{ref}}\right)^2 \quad (13)$$

where $Mark_{TA}$ denotes the mark related to tracking effectiveness, $f_{ref}$ represents an arbitrary reference frequency, $MarkRef_{TA}$ is the corresponding reference mark, and $m_{TA}$ is a proportionality constant.

Regarding Doppler spread and channel coherence time, either if we consider scalable subcarrier widths or not, their individual effects would lead to the same type of linear dependency with frequency as tracking effectiveness has (in the log domain):

$$Mark_D = MarkRef_D - m_D \cdot log\left(\frac{f}{f_{ref}}\right) \quad (14)$$

$$Mark_{Coh} = MarkRef_{Coh} - m_{Coh} \cdot log\left(\frac{f}{f_{ref}}\right) \quad (15)$$

Absorbing the different factors and constants we would then have an overall linear dependency of the mark in the log domain as follows:

$$Mark = Mark_{TA} + Mark_D + Mark_{Coh} = MarkRef - m \cdot log\left(\frac{f}{f_{ref}}\right) \quad (16)$$

where $m \equiv 2m_{TA} + m_D + m_{Coh}$. The factor 2 accounts for the quadratic dependence of tracking effectiveness with frequency. In case of considering ideally scalable subcarrier widths, $m_D$ would equate to zero thus making the system insensitive to Doppler spread. As noted above, perhaps practical systems could only afford this at spot frequencies and dependence of the mark would still be linear out of those points.

The two combined parameters *MarkRef* and *m* can be adjusted so as to reflect the relative differences in mobility support across different frequencies. Assigning values for the co-efficients, the assignment of $m_{TA}$, $m_D$ and $m_{Coh}$ is somewhat arbitrary, only the relative variations make sense when comparing different frequencies. Successful field trials were conducted at 28 GHz by Samsung reaching 1.2 Gbps with 100 km/h mobility [10]. Nokia and DoCoMo are also planning to extend their current 70 GHz indoor trials to outdoors, initially with pedestrian mobility [11]. While the beam tracking problem is quite novel to the industry, as even in 4G, most of the coverage provided is with 3-sector or even omni directional cells. Thus beam tracking issue can be considered as more challenging to the industry. Reflecting the above facts, we propose to induce double the value of $m_D$ and $m_{Coh}$ to $m_{TA}$. The suggested values are $m_{TA}=2$ and $m_D$ and $m_{Coh}=1$.

E. *Application of the figure of merit for Strand 1*

The figure of merit obtained above can be thus applied in its most general form, with all the three mentioned effects: tracking effectiveness, Doppler spread, and channel coherence time. Their combined effects can be absorbed into suitable parameters $MarkRef|_{BW}$ and $m|_{BW}$ that should be adjusted to reflect mobility support across frequency:

$$Mark = MarkRef|_{BW} - m|_{BW} \cdot log\left(\frac{f}{f_{ref}}\right) \quad (17)$$

where $m|_{BW} \equiv 2m_{TA} + m_D + m_{Coh}$.

With the assigned values as stated above, the combined value $m|_{BW}=6$ is applied for strand 1. $MarkRef|_{BW}$ value is taken as 10, so at the reference frequency (6 GHz), the FoM is scaled to 10. Although continuous values could be obtained for FoM on mobility, we only depict the spot frequency values as in line with the other KPIs.

| Carrier frq (GHz) | 6 ($f_{ref}$) | 28 | 38 | 73 |
|---|---|---|---|---|
| Figure of Merit | 10 | 5.99 | 5.19 | 3.49 |

Table 5: Mobility KPI analysis – Strand 1

F. *Application of the figure of merit for the Strand 2*

Strand 2 lets the bandwidth change with frequency while the number of antennas is kept fixed (e.g. 32). Tracking effectiveness is therefore constant with frequency and the only effects are Doppler spread and channel coherence time, which lead to a similar linear dependence:

$$Mark = MarkRef|_{no.antennas} - m|_{no.antennas} \cdot log\left(\frac{f}{f_{ref}}\right) \quad (18)$$

where in this case $m|_{no.antennas} \equiv m_D + m_{Coh}$. Bandwidth itself has no effect on mobility, but the implicit frequency variations will impact mobility through the two above mentioned effects. With the assigned co-efficient values, the combined co-efficient $m|_{no.antennas}=2$, for strand 2.

Table 6: Mobility KPI analysis – Strand 2

| Carrier frq (GHz) | 6 ($f_{ref}$) | 28 | 38 | 73 |
|---|---|---|---|---|
| Figure of Merit | 10 | 8.66 | 8.4 | 7.83 |

## VI. Conclusions

In this paper, we have analyzed the behavior of the millimetre-wave spectrum under the 3 key cellular KPIs – coverage, capacity and mobility. The overall analysis is a highly complex process, so we attempt to break down the problem into 2 strands – analysis under incrementing antenna numbers with fixed bandwidth and analysis under incrementing bandwidth with fixed antenna numbers. We achieve a single Figure of Merit (FoM) for each of the strands for each of the KPI's. We also note the significant limitations in some of our assumptions. The coverage and capacity FoM evaluation is conducted through simulations, while the mobility FoM evaluation is analytical.

The results indicate that all FoMs for all 3 KPIs degrade under increasing frequency in this millimeter-wave spectrum, for both strands considered. The coverage and capacity analysis is based on spot frequency simulations and hence show some sharp changes in the FoM. The main take away from this study should be the trends in the changing FoMs and not the absolute values of FoM – as many complex factors, which may be overlooked in this analysis, can influence the precise FoM values. The trends indicate the adaptation of higher millimeter-wave spectrum is more challenging for cellular systems, but these challenges can be overcome.

## Acknowledgements

The European Commission funding under H2020- ICT-14-2014 (Advanced 5G Network Infrastructure for the Future Internet, 5G PPP), and project partners: Samsung, Ericsson, Aalto University, Alcatel-Lucent, CEA LETI, Fraunhofer HHI, Huawei, Intel, IMDEA Networks, Nokia, Orange, Telefonica, Bristol University, Qamcom, Chalmers University of Technology, Keysight Technologies, Rohde & Schwarz, TU Dresden are acknowledged.